# Nitrogen-Nitrogen Bonds Violate Stability of N-Doped Graphene

Vitaly V. Chaban[1,2] and Oleg V. Prezhdo[2]

1) Instituto de Ciência e Tecnologia, Universidade Federal de São Paulo, 12231-280, São José dos Campos, SP, Brazil

2) Department of Chemistry, University of Southern California, Los Angeles, CA 90089, United States

**Abstract**. Two-dimensional alloys of carbon and nitrogen represent an urgent interest due to prospective applications in nanomechanical and optoelectronic devices. Stability of these chemical structures must be understood as a function of their composition. The present study employs hybrid density functional theory and reactive molecular dynamics simulations to get insights regarding how many nitrogen atoms can be incorporated into the graphene sheet without destroying it. We conclude that (1) C:N=56:28 structure and all nitrogen-poorer structures maintain stability at 1000 K; (2) stability suffers from N-N bonds; (3) distribution of electron density heavily depends on the structural pattern in the N-doped graphene. Our calculations support experimental efforts on the production of highly N-doped graphene and tuning mechanical and optoelectronic properties of graphene.

TOC Image

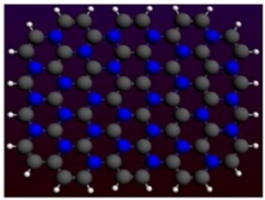 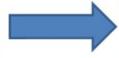 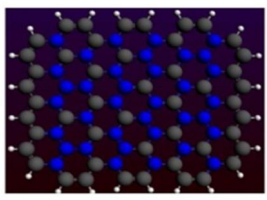

**Introduction**

Graphene is a carbonaceous nanomaterial with scientific impact and technological promise. The graphene field of research currently witnesses a variety of novel synthesis approaches and applications.[1-28] Being extremely chemical versatile, carbon atoms can link with one another tetrahedrally to form diamond. Otherwise, they can arrange in layers, with a genuine chicken-wire structure, to produce graphite. The hexagonally arranged carbon-carbon bonds are rich in high-energy electrons. This important feature makes graphite to conduct electricity along the layer. Graphene consists of just a single layer of graphite. Various morphologies of graphene have also been obtained including two-dimensional graphene nanosheets,[27] one-dimensional graphene nanoribbons,[29] and zero-dimensional graphene quantum dots.[30]

The electronic properties of graphene are quite different from those of other novel carbonaceous nanostructures, such as graphite, fullerenes and nanotubes.[2,14,27,31-33] The properties of nanoribbons and quantum dots can be efficiently tuned by adjusting their size and chemical composition of edges. For instance, very narrow nanoribbons (< 10 nm) exhibit semiconducting behavior, whereas wider nanoribbons exhibit very weak gate dependence. The zigzag edge decreases a band gap.

The charge carriers in graphene are unusual, since they originate from interactions within the symmetrical periodic arrangement of electrons spread across the hexagonal lattice. The lattice is responsible for waves of electric charge. These are known as quasi-particles. Such quasi-particles behave like photons and retain the quantum features of electrons (charge and spin). The electronic charge carriers can be described by the special theory of relativity. This opens avenues to test the relativistic theory using graphene and, to certain extent, avoid high-energy particle accelerators.

Not only conformations of graphene are important to adjust its properties, the electronic and nanomechanical properties can also be tuned by doping other elements.[14,22,30,34-39] Nitrogen, boron, sulfur and silicon can be readily incorporated into the hexagonal lattice of graphene without breaking its overall structure.[39,40] By altering chemical, mechanical and electrical properties of graphene, these dopants open a myriad of possibilities[34-39] to extend existing applications of pure graphene and the two-dimensional boron nitride.

The N-doped graphene exhibits drastically different properties as compared to the pristine graphene.[5,14,22,30,34,35,37,38,40] The spin and electron density distributions for carbon atoms are influenced by the neighbor nitrogen add-atoms, which induce the activated region on the graphene surface. Such an activated region participates in the catalytic reactions directly. For instance, it fosters the oxygen reduction reaction and anchors the metal nanoparticles used in the catalytic reaction. N-doping, in addition, shifts the Fermi level above the Dirac point. Consequently, the density of state near the Fermi level is suppressed. The band gap between the conduction band and the valence band gets opened. The possibility to engineer a band gap for graphene makes it a suitable candidate for semiconductor devices. N-graphene is also an interesting compound in the context of sensors, batteries, and supercapacitors. N-doping of graphene greatly broadens its laboratory and industrial applications.[22,30,34,35,40]

The review paper of Liu et al.[41] focuses on the two chemical doping approaches and band gap tuning. In turn, the recent review paper of Wang and coworkers[39] summarizes different synthesis and characterization methods of N-substituted graphene and discuss potential applications of N-graphene based on the existing experimental and theoretical studies. It has been reported that $C_3N_4$ and $C_6N_9H_3$ compositions of graphene feature a large band gap, ca. 5 eV.[42] The band gap can be tuned by external stress or add-atoms. In the case of the delta-doped graphene, the band gap is opened only when the content of nitrogen exceeds 25%. It was also noted[43] that delta-doping is not a likely substitution product in real chemical practice. When the nitrogen concentration falls between 2-4%, the decay of the diffusion coefficient is observed. This indicates the onset of quantum interference effects.[44] As long as the nitrogen content is low, this effect is weak and affects conduction only marginally. While maintaining competitive mobility and conductivity, N-doping modulates graphene properties. Unfortunately, the real-world doping of graphene results not only in the introduction of the nitrogen atoms, but also in the introduction of defects. Defects constitute scattering centers affecting mobility of graphene. When the nitrogen concentration rises above 5%, a strong localization effect may be expected, which will decrease the mobility in the N-graphene.[39]

This work reports a combined simulation research employing long-time reactive molecular dynamics (RMD) simulations[45] and hybrid density functional theory (HDFT) to understand stability of the N-doped graphene sheets as a function of their chemical composition and internal structure. We consider five different cases of N-doped graphene and test their sustainability upon thermal motion at 1000 K. Partial electrostatic charges localize on every atom and molecular orbitals in the valence and conduction band are provided and discussed. Electric structure analysis is performed for geometrically optimized molecular configurations. We conclude that the most vulnerable structural pattern is a nitrogen-nitrogen bond, whereas carbon-nitrogen bonds remain perfectly stable over the course of high-temperature RMD simulations.

**Computational Methodology**

Five major systems standing for different compositions of the N-doped graphene (Table 1) are considered in this work. These particular systems were constructed to test a few different arrangements of the host atoms and dopants upon increasing nitrogen:carbon ratios.

Table 1. Simulated systems: their properties and stability. The stability was identified based on the 10 ns long reactive molecular dynamics runs at 1000 K. Note that percentage of nitrogen atoms is provided with respect to all atoms in the hexagonal N-doped graphene sheet. This detail must be kept in mind when comparing our stability results with other experimental and theoretical investigations

| # | Formula | # electrons | stability | C:N ratio | N (%) | Comments |
|---|---------|-------------|-----------|-----------|-------|----------|
| I | $C_{74}N_{10}H_{24}$ | 538 | stable | 74:10 | 9.3 | 1 N per aromatic ring |
| II | $C_{64}N_{20}H_{24}$ | 548 | stable | 64:20 | 18.5 | 2 N per aromatic ring |
| III | $C_{51}N_{33}H_{24}$ | 560 | unstable | 51:33 | 30.6 | 3 N per aromatic ring |
| IV | $C_{56}N_{28}H_{24}$ | 556 | stable | 56:28 | 25.9 | max. N without N-N bonds |
| V | $C_{48}N_{36}H_{24}$ | 564 | unstable | 48:36 | 33.3 | 4 N per aromatic ring |

The molecular dynamics (MD) simulations were performed using quantum chemistry (QC) based reactive force field (ReaxFF).[45-47] This methodology was applied previously with success to address a number of complicated problems in organic chemistry and materials science. ReaxFF provides a nearly ab initio level of description of reactive potential surfaces for many-particle systems.[45-47] The method treats all atoms in the system as separate interaction centers. The instantaneous point charge on each atom is determined by the electrostatic field due to all surrounding charges, supplemented by the second-order description of dE/dq, where E is internal energy and q is electrostatic charge on a given atom. The interaction between two charges is written as a shielded Coulomb potential to guarantee correct behavior of covalently bonded atoms. The instantaneous valence force and interaction energy between two atoms are determined by the instantaneous bond order. The latter is determined by the instantaneous bond distance. These interaction energy functions are parametrized vs. QC energy scans involving all applicable types of bond-breaking processes. The bond order concept is used to define other valence interactions, such as bond, lone electron pair, valence angle, conjugation, and torsion angle energies. It is important for energy conservation and stability that all interaction terms smoothly decay to zero during bond dissociation. The conventional pairwise van der Waals energy term describes short-range electron-electron repulsion, preserving atom size, and longer-range London attractive dispersion. Unlike non-reactive MD simulations, ReaxFF uses the van der Waals term for covalently bonded atoms, where it competes with a monotonically attractive bond term. Such an approach to chemical bonding requires a significant number of independent parameters, which can be obtained from QC energies.[45-47] Bond dissociation, geometry distortion, electrostatic charges, infrared spectra, equations of state and condensed-phase structure are typically derived using an electronic structure method, such as density functional theory, to be consequently used in the ReaxFF parametrization. The works by van Duin, Goddard and coworkers[45-47] provide a more comprehensive description of the methodology used here. The implementation in ADF2013 (scm.com) was used for these simulations.

The HDFT calculations on the doped graphene sheets were performed using the recently proposed functional wB97xD.[48] This functional performs well for electronic structure, thermochemistry and inter-molecular binding energies. It also includes an empirical correction for dispersive attraction. Due to its good reputation, wB97xD is preferred over more traditional HDFT approaches. The wave function was expanded using the Gaussian 6-311G basis set supplemented by polarization and diffuse functions on every non-hydrogen atom, 6-311+G*. This basis set provides a reasonable balance between an affordable computational cost for a relatively large system and accuracy of results (electronic energy levels, partial electrostatic charges, molecular orbitals).[48] All electrons in the systems were represented explicitly requiring no effective-core potentials. Prior to calculation properties, the geometries of the N-doped graphene sheets were submitted to internal energy minimization by the conjugate gradient algorithm. Frequency analysis which followed after each geometry optimization ensured absence of negative modes. The implementation in Gaussian 09, rev. D (www.gaussian.com) was used for these calculations.

**Results and Discussion**

Figure 1 summarizes initial geometries of the N-doped graphene sheets. System I features randomly substituted carbon atoms by nitrogen atoms under the condition that no more than a single nitrogen atom occurs per carbon ring. The same conditions are fulfilled for systems II and III, where no more than two and three nitrogen atoms per six-membered ring were allowed. System IV contains maximum possible number of nitrogen atoms, but without any nitrogen-nitrogen bond. Finally, system V contains four nitrogen atoms per ring. Note that no boundary nitrogen atom was substituted, i.e. there is no nitrogen-hydrogen bond.

Reactive molecular dynamics simulations[46] provide an efficient and relatively inexpensive tool to observe stability of the constructed chemical structures at finite temperature (Figures 2-3). Systems I-II appear perfectly stable during the 10 ns long RMD simulations at 1000 K. However, systems III and V are less stable. System V decomposes during the first nanosecond of the RMD simulation. System III, in turn, destructs a number of its bonds, mostly nitrogen-nitrogen bonds, but preserves certain structure during the entire 10 ns long RMD. System IV does not contain nitrogen-nitrogen bonds, but consists of 25.9% nitrogen atoms (with respect to all other atoms including hydrogen atoms). This doped structure maintains stability, in contrast to system III where many nitrogen-nitrogen bonds disappear readily.

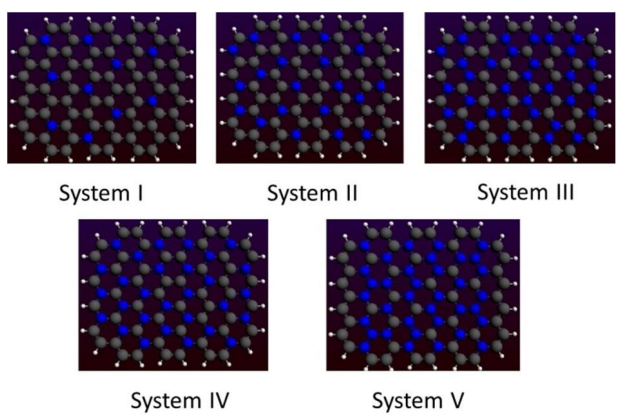

Figure 1. The initial geometries of the five investigated systems, see Table 1.

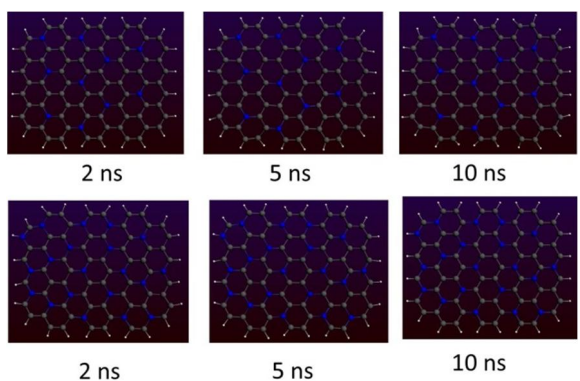

Figure 2. The atomistic-precision snapshots taken in the course of reactive molecular dynamics run: systems I (top) and II (bottom).

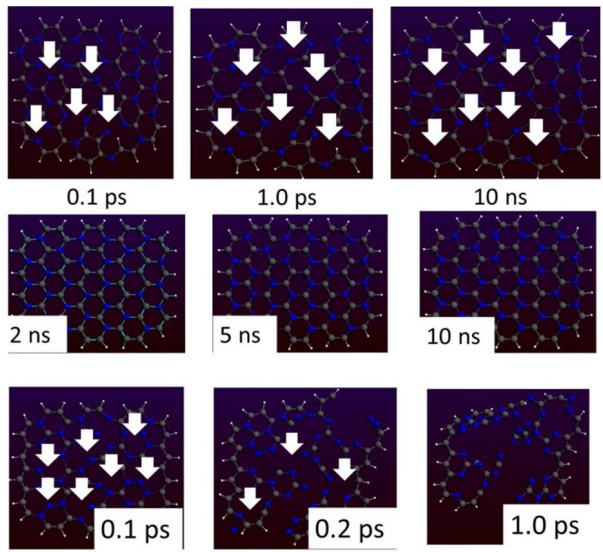

Figure 3. The atomistic-precision snapshots taken in the course of reactive molecular dynamics run: systems III (top), IV (middle) and V (bottom). The white arrows point disruptions of the N-graphene structures. Note the most broken covalent bonds were N-N bonds.

Figures 4-7 provide a detailed description of the N-doped graphene structures in terms of carbon-carbon, carbon-nitrogen and nitrogen-nitrogen radial distribution functions. The carbon-carbon covalent bond length amounts to ca. 0.14 nm and fluctuates insignificantly, due to its strength and aromatic nature. This bond remains stable during molecular dynamics at 1000 K in most considered cases (Table 1). Nitrogen doping does not influence either length or stability of this bond. Consider multiple peaks at 0.14, 0.25, 0.29, 0.38, 0.43, 0.50, 0.52, 0.58, 0.63, 0.66 nm (system I). Out of these, only the peak at 0.14 nm corresponds to the covalent C-C bonds, whereas all other sharp and relatively high peaks arise due to a distinct arrangement of carbon atoms in the hexagonal cells. These peaks must provide an important support for the interpretation of the experimental X-ray data for N-doped graphene.

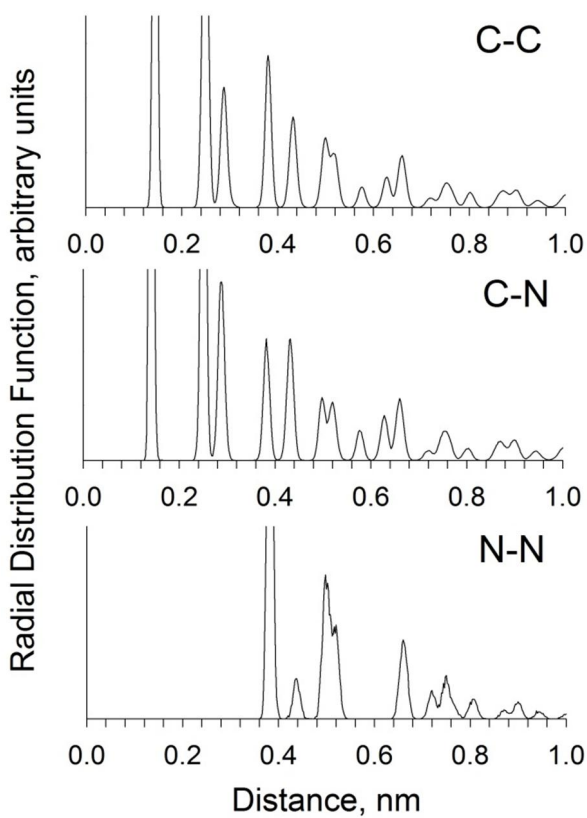

Figure 4. Radial distribution functions for the carbon-carbon, carbon-nitrogen, and nitrogen-nitrogen atomic pairs in system I, see Table 1. The functions were evaluated following 10 ns long reactive molecular dynamics simulations at 1000 K.

The carbon-nitrogen bond also exhibits a decent stability (Figures 4-7), being of similar length as the C-C covalent bond. The sharp and high peaks indicate that oscillations of this bond upon equilibrium dynamics are small. Breakage of this bond is unlikely if the simulation of finite-temperature dynamics is extended to greater time scales. Not only the covalent carbon-nitrogen distances are well defined, but also longer (non-covalent) distances are distinguished well. Consider peaks at 0.14, 0.25, 0.29, 0.38, 0.43, 0.50, 0.52 nm, etc (system II). The discussed correlations persist during ca. 1.0 nm. Note a clear similarity between the C-C and C-N peak locations. Carbon and nitrogen atoms possess nearly the same van der Waals size, which allows them to conveniently integrate into the crystalline cells of one another with minor perturbations of the overall structures. If only this point of reference of considered, graphene can be infinitely doped by nitrogen atoms. However, we know that pure nitrogen does not produce rings and thus the N-doped grapheme sheet must transform into molecular nitrogen at certain dopant concentration. The goal of the present study is to provide RMD and electronic structure evidence regarding sustainability of the N-doped graphene sheets.

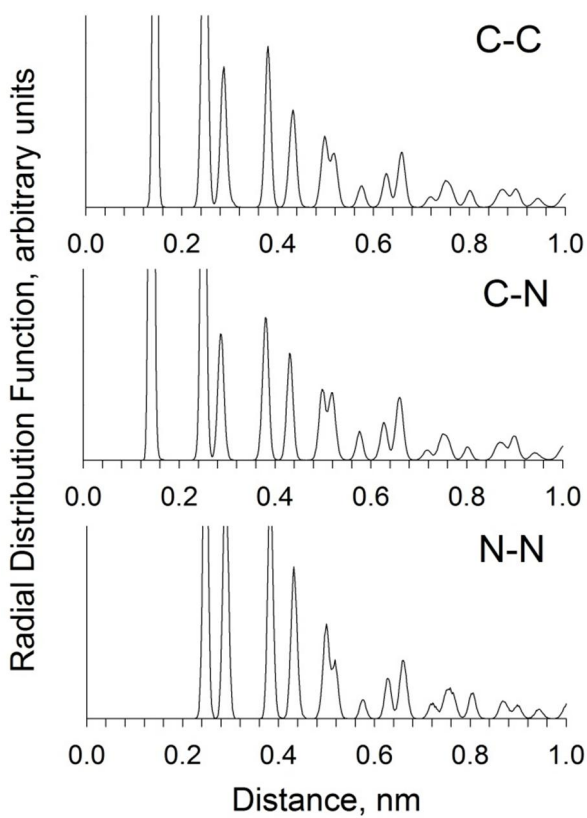

Figure 5. Radial distribution functions for the carbon-carbon, carbon-nitrogen, and nitrogen-nitrogen atomic pairs in system II, see Table 1. The functions were evaluated following 10 ns long reactive molecular dynamics simulations at 1000 K.

On the contrary, nitrogen-nitrogen bonds, systems III and V, exhibit poorer stability and break apart during the first hundreds and thousands of molecular dynamics steps. While in system III, destruction of the N-N bonds leads to formation of the nine- and more-membered rings (Figure 3), which remain relatively stable, system V loses its graphene-like shape very quickly. System IV preserves stability thanks to an even distribution of nitrogen atoms throughout the surface despite comparable molar fraction of the dopant. Recall than system IV does not contain N-N bonds, but contains abundant C-N bonds. Compare RDF peaks at 0.25, 0.29, 0.38, 0.42, 0.49, 0.64 nm (system III) and at 0.12, 0.15, 0.24, 0.43 nm (system V). The total number of N-N RDF peaks is significantly smaller than the total number of either C-C or C-N peaks in any system. A great advantage of RMD is its quite realistic description of bonded parameters (lengths, angles) and their dynamics in the course of equations-of-motion propagation. Usage of simpler, pairwise interaction potentials provides too simplified description, which is unacceptable for the present highly precise study and for comparison with the experimental data in the case of non-trivial chemical structures.

If atomistic oxygen in excess is supplied to the systems, it will foster oxidation of system V towards $CO_2$, $NO_2$ gases, and $H_2O$ vapor. If a lack of oxygen is provided, nitrogen may remain in the $N_2$ state depending on temperature and other reactants. Note that the absence of covalent bonding peaks in Figures 4-5 corresponds to an initial absence of the N-N bonds in these structures, rather than to their breakage during molecular dynamics. Sharp non-covalent peaks are present in these systems indicating integrity of the N-doped graphene sheet.

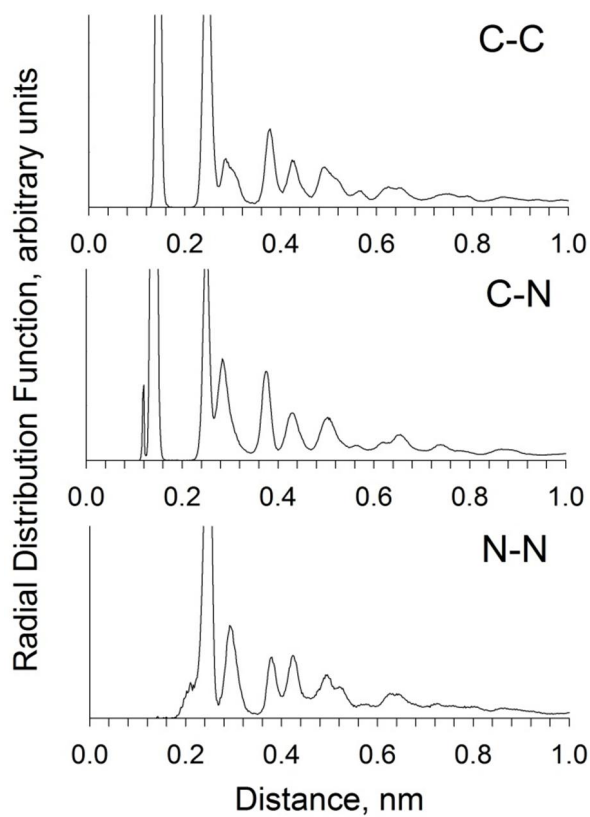

Figure 6. Radial distribution functions for the carbon-carbon, carbon-nitrogen, and nitrogen-nitrogen atomic pairs in system III, see Table 1. The functions were evaluated following 10 ns long reactive molecular dynamics simulations at 1000 K.

Interestingly, the marginal peak at ca. 0.17 nm (Figure 7) is still present corresponding to sporadic N-N bonds, although the molecular snapshots clearly indicate collapse of the initial structure. Furthermore, the observed N-N bond length is much larger than that of conventional N-N bond, 0.12-0.13 nm upon high-temperature thermal motion. As opposed to conclusions of Figures 4-5, the longer-range N-N structure is absent in system V.

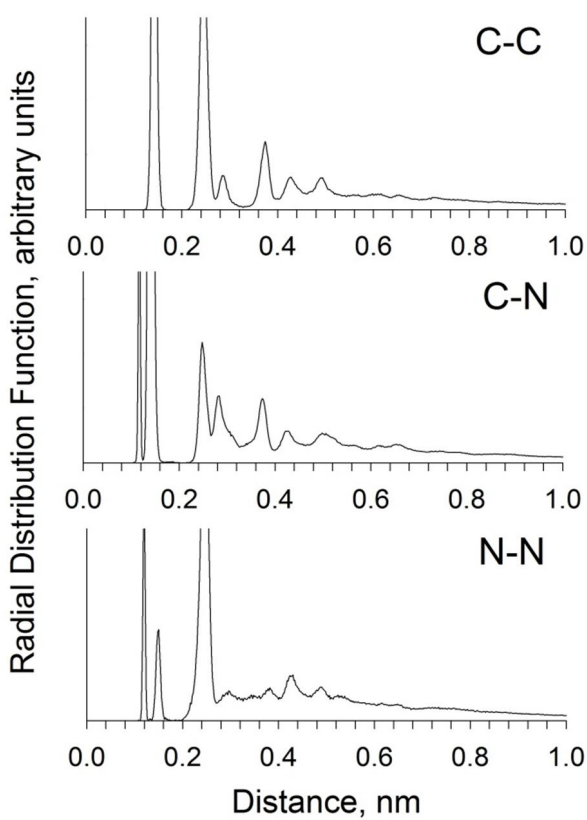

Figure 7. Radial distribution functions for the carbon-carbon, carbon-nitrogen, and nitrogen-nitrogen atomic pairs in system V, see Table 1. The functions were evaluated following 10 ns long reactive molecular dynamics simulations at 1000 K.

Figure 8 summarizes valence and conduction band molecular orbitals (MOs) for the four N-doped graphene sheets. These orbitals had been computed prior to molecular dynamics at 1000 K. They correspond to the optimized geometry configurations. System III can be optimized into a stable geometry; however, it breaks apart at 1000 K. All depicted MOs are preferentially localized on the carbon atoms, rather than on hydrogen and nitrogen atoms. The graphene sheets exhibiting less stability, in general, have somewhat more localized MOs, although the difference is not drastic. The highest MO of the nitrogen atoms lies lower in the energy diagram then HOMO of sheet. We anticipate that nitrogen atoms significantly alter a valence electronic band of pristine graphene, which is a primary reason for instability upon the critical dopant content.

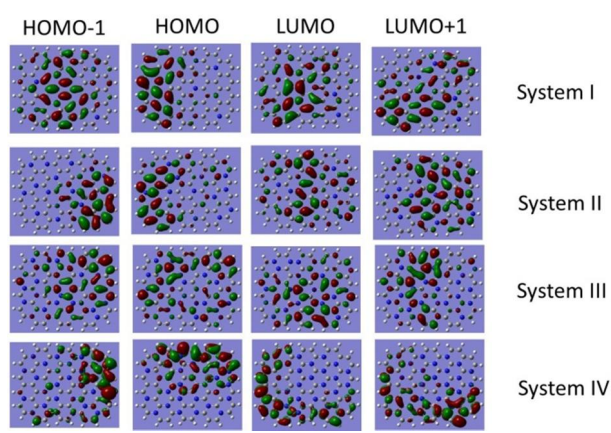

Figure 8. Visualization of the selected molecular orbitals (HOMO-1, HOMO, LUMO, and LUMO+1) in systems I, II, III, IV. The calculations are based on the wB97xd/6-311+G(d) level of theory. The geometries correspond to their local minima achieved by the conjugate gradient internal energy minimization algorithm. These geometries do not account for thermal motion. Note that system III, which is unstable at 1000 K, preserved its geometry at 0 K.

Atomic charges on carbon, nitrogen and hydrogen atoms supplement the outlined observations from RMD and molecular orbitals and provide additional description of the N-doped graphene sheet instability upon an excessive content of the dopant. Figures 9-11 provide Hirshfeld charges on every atom of systems I, III, IV. These particular systems were chosen, since they preserve stability after geometry optimization, despite N-doping. System II is omitted since its behavior is similar to that of system I. Figure 12 summarizes observations by depicting total charges on C, N, and H atoms in the corresponding systems. Certain similarities of all stable graphene sheets should be outlined.

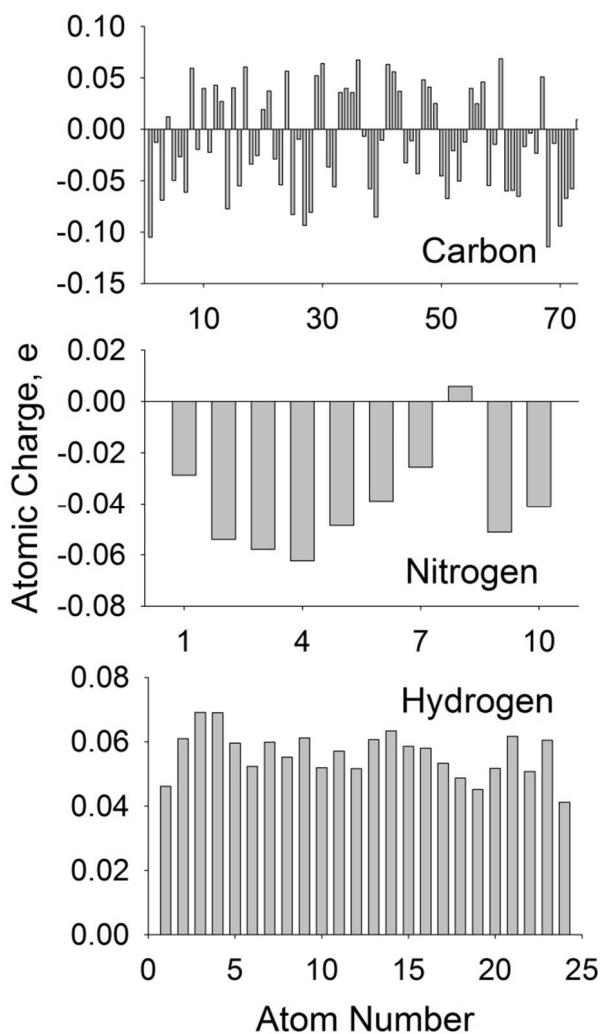

Figure 9. The atomic charges (in electron units) localized on carbon, nitrogen, and hydrogen atoms of system I. The atomic charges were defined according to the well-known Hirshfeld scheme. The calculated atomic charges correspond to the local minima of the corresponding geometries achieved at the wB97xd/6-311+G(d) level of theory. The depicted atom numbers refer to serial numbers of the corresponding atoms in the simulated systems and do not reflect chemical structures.

First, carbon atoms attain both positive and negative electron density. The total charge on the carbon atoms in system III is nearly. Second, all hydrogen atoms attain a small positive charge, usually less than +0.1e. Third, all nitrogen atoms are negatively charged with only single exception in systems I and III and no exception in system IV. Therefore, larger content of nitrogen atoms favor accumulation of electrons on them. Although the accumulated charges are relatively small, systematic electron accumulation may on nitrogen, instead of carbon, leads to polarization of carbon-nitrogen covalent bonds. This may results in making them weaker.

Additionally, N-N bonds are unstable due to excess of electron density on both N atoms participating in the covalent bonding. Accumulation of negative charge on nitrogen atoms upon their stepwise addition to the sheet is clearly seen in Figure 12.

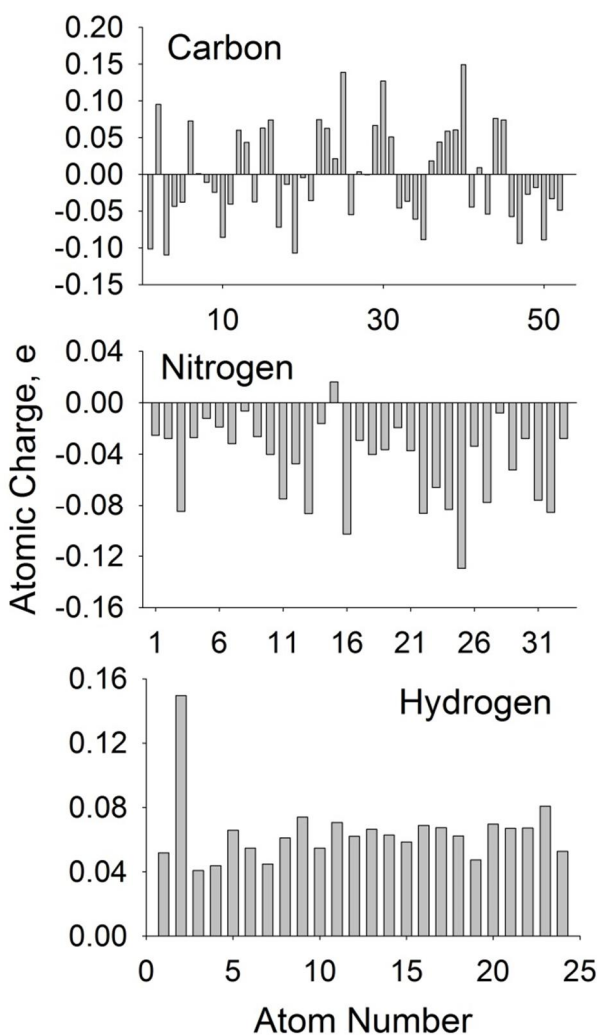

Figure 10. The atomic charges (in electron units) localized on carbon, nitrogen, and hydrogen atoms of system III. The atomic charges were defined according to the well-known Hirshfeld scheme. The calculated atomic charges correspond to the local minima of the corresponding geometries achieved at the wB97xd/6-311+G(d) level of theory. The depicted atom numbers refer to serial numbers of the corresponding atoms in the simulated systems and do not reflect chemical structures.

Having conducted RMD finite-temperature simulations, molecular orbital analysis based on the hybrid DFT and electron density localization analysis, we conclude that the major vulnerability of the N-doped graphene sheet is linked to the presence of the N-N bonds. The N-doped graphene sheet is stable irrespective of the nitrogen content provided that only C-C and C-N covalent bonds are present. In turn, the N-N bonds break apart during the very first steps of RMD giving rise to nine- and more-membered rings. These rings lose their aromaticity and form energetically unfavorable dangling bonds. These bonds cannot be saturated in the absence of hydrogen source. Liberated higher-energy p-electrons ionize carbon and nitrogen atoms. Thus, the second step is the destruction of carbon-nitrogen bonds. The decomposition of the N-doped graphene terminates at this stage, as no additional reactants were provided. Upon real conditions, the reactions must go further to form simple and stable monomolecular products.

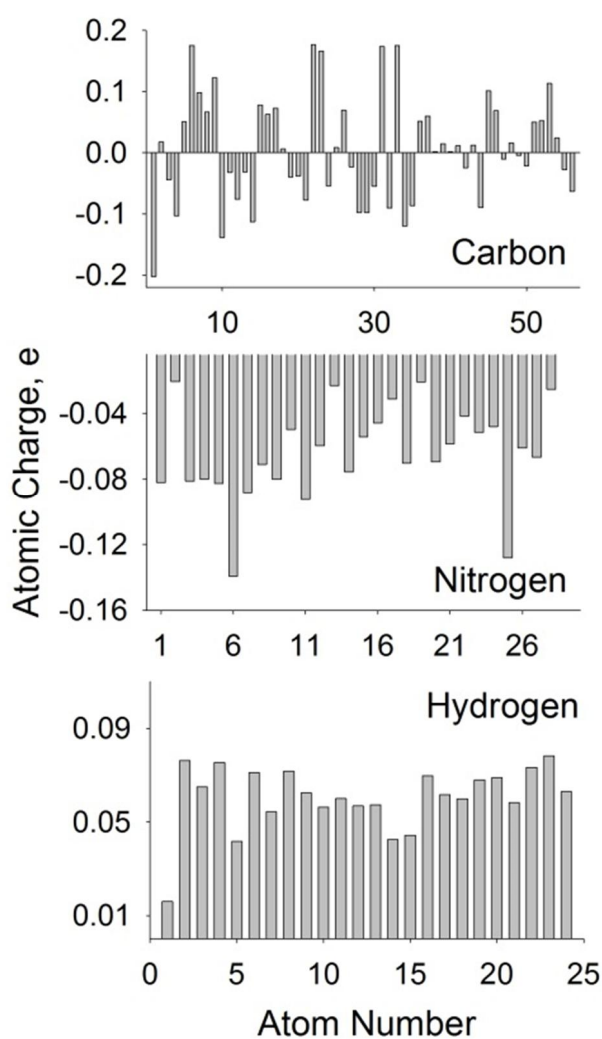

Figure 11. The atomic charges (in electron units) localized on carbon, nitrogen, and hydrogen atoms of system IV. The atomic charges were defined according to the well-known Hirshfeld scheme. The calculated atomic charges correspond to the local minima of the corresponding geometries achieved at the wB97xd/6-311+G(d) level of theory. The depicted atom numbers

refer to serial numbers of the corresponding atoms in the simulated systems and do not reflect chemical structures.

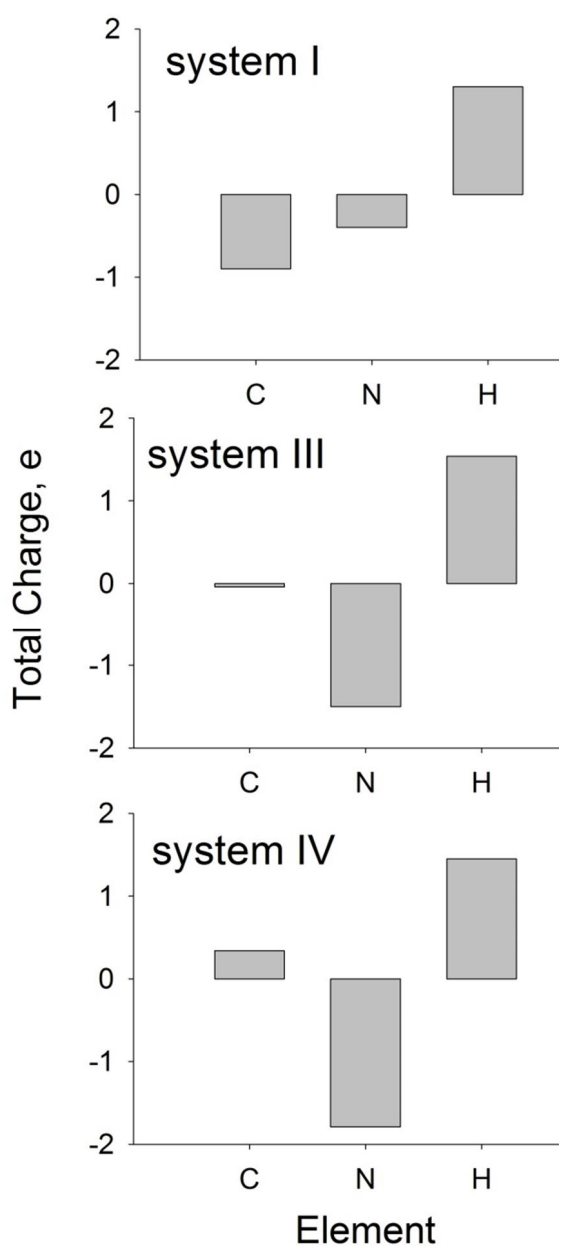

Figure 12. The atomic charges (in electron units) localized on carbon, nitrogen, and hydrogen atoms of system III. The atomic charges were defined according to the well-known Hirshfeld scheme. The calculated atomic charges correspond to the local minima of the corresponding geometries achieved at the wB97xd/6-311+G(d) level of theory. The depicted atom numbers refer to serial numbers of the corresponding atoms in the simulated systems and do not reflect chemical structures.

**Conclusions**

The reported study shows that N-doped graphene[5,14,22,30,34,35,37-41] can accumulate nitrogen atoms until N-N bonds are formed, in addition to C-N bonds. The latter are observed to be stable at 1000 K. Obviously, higher-temperature dynamics and especially the presence of various structure defects can break these bonds. In turn, the fused six-membered rings containing N-N bonds are not stable, according to our simulations. They ruin during the first molecular dynamics iterations, re-arranging into nine- and more-membered rings. The latter are vulnerable to oxidation and other chemical reactions, which would result in stable low-molecular products. Our results provide guidance for engineering N-doped graphene sheets with high nitrogen content, as nitrogen doping constitute efficient means to tune, adjust and refine mechanical and electrical properties of graphene.

**Acknowledgments**

This research was in part supported by grant CHE-1300118 from the US National Science Foundation. V.V.C. is a CAPES fellow under the "Science Without Borders" program.